# A Tale of Two Sides of Wafer: Physical Implementation and Block-Level PPA on Flip FET with Dual-sided Signals

Haoran Lu, Xun Jiang, Yanbang Chu, Ziqiao Xu, Rui Guo, Wanyue Peng, Yibo Lin, Runsheng Wang, Heng Wu†, Ru Huang

School of Integrated Circuits, Peking University, Beijing 100871, China

†Email: hengwu@pku.edu.cn

*Abstract*—As the conventional scaling of logic devices comes to an end, functional wafer backside and 3D transistor stacking are consensus for next-generation logic technology, offering considerable design space extension for powers, signals or even devices on the wafer backside. The Flip FET (FFET), a novel transistor architecture combining 3D transistor stacking and fully functional wafer backside, was recently proposed. With symmetric dual-sided standard cell design, the FFET can deliver around 12.5% cell area scaling and faster but more energy-efficient libraries beyond other stacked transistor technologies such as Complementary FET (CFET). Besides, thanks to the novel cell design with dual-sided pins, the FFET supports dual-sided signal routing, delivering better routability and larger backside design space. In this work, we demonstrated a comprehensive FFET evaluation framework considering physical implementation and block-level power-performance-area (PPA) assessment for the first time, in which key functions are dual-sided routing and dual-sided RC extraction. A 32-bit RISC-V core was used for the evaluation here. Compared to the CFET with single-sided signals, the FFET with single-sided signals (for fair comparison) achieved 23.3% post-P&R core area reduction, 25.0% higher frequency and 11.9% lower power at the same utilization, and 16.0 % higher frequency at the same core area. Meanwhile, the FFET supports dual-sided signals, which can further benefit more from flexible allocation of cell input pins on both sides. By optimizing the input pin density and BEOL routing layer number on each side, 10.6% frequency gain was realized without power degradation compared to the one with single-sided signal routing. Moreover, the routability and power efficiency of FFET barely degrades even with the routing layer number reduced from 12 to 5 on each side, validating the great space for cost-friendly design enabled by FFET.

## I. Introduction

With the ending of conventional scaling of logic transistors, the advanced logic technology development follows two general trends. One is the functional wafer backside, starting from backside power delivery network (BSPDN) [1], [2], [3], [4], [5] and backside signals [3], [4], [5], [6], [7] and moving towards backside devices [3]. The other one is the advanced device architecture, trending towards 3D transistor tacking. Without 3D transistor stacking, the scaling of standard cell height ends at around 6-track (6T, 1T defined as 1 M2 pitch) FinFET or 5T Nanosheet FET. While, stacked transistor can easily break this limitation. The Complementary FET (CFET) [8], composed of a nFET stacked vertically over a pFET, was proposed as an emerging solution to the ultra-scaled 3D transistor stacking, with great reduction in cell footprint. The standard cell height of CFET is further compressed by using buried power rail (BPR) on the wafer backside for reduced area and better performance. This novel architecture of CFET results in more scaled cell designs of 4T [8], [9] or even 3T [8], [10] height.

However, with the ultimate standard cell size scaling, routing is undoubtedly more complicated and should be carefully examined. Backside signal routing is one of the key enablers to boost routability [4]. Unfortunately, due to the limitation of device structure and process complexity, it almost impossible to design standard cells with pins inherently on the wafer backside no matter whether using FinFET, Nanosheet FET or CFET. In order to realize backside signal routing, signals have to be transferred between frontside and backside through bridging cells [4], [7] (e.g. buffers), which will lead to significant area penalty and design complexity. Specifically, CFET's ultra-scaled standard cell design features very high pin density, thus worse routability [11].

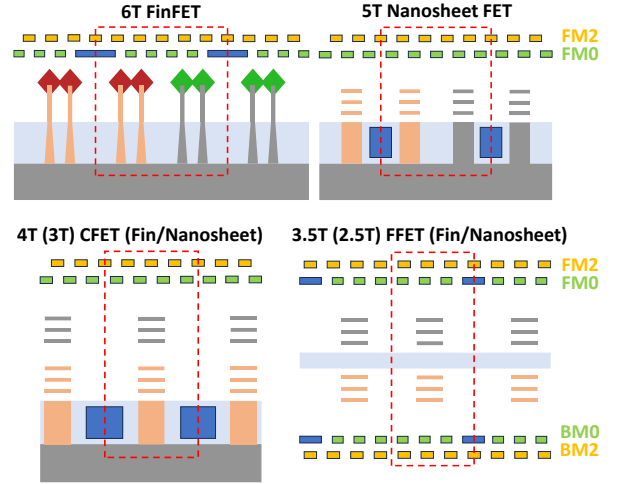

Fig. 1. Scaling roadmap from FinFET to Nanosheet FET, CFET and FFET.

As a solution, H. Lu et al. [12] proposed a novel transistor architecture called Flip FET (FFET), which is named from the wafer flipping process, combing 3D transistor stacking and fully functional backside. Different from the CFET, the FFET is consisted of a pair of nFET and pFET which are back-to-back-stacked on each other. As the backside transistor is formed after the frontside process through a similar backside process flow, FEOL and BEOL structures are almost identical on both sides in the FFET architecture. This enables the unique symmetric standard cell design (intra-cell signals and power rails routed symmetrically on both sides) with pins inherently set on the backside, facilitating much easier backside signal routing in the FFET architecture. Due to this symmetric cell design, the FFET can achieve further area scaling beyond the CFET in the roadmap, with 3.5T or even 2.5T height as shown in Fig. 1. However, there is still lack of a powerplan scheme for the symmetric dual-sided power rails in the FFET. Moreover, although the concept of dual-sided signal routing was mentioned earlier [12], the specific implementation algorithm was not disclosed while current P&R tools do not support cells with pins inherently located on the backside. Last but not least, whether the FFET with dual-sided routing can truly demonstrate advantages over the CFET at the block level has yet to be proven.

In this paper, we propose a comprehensive FFET evaluation framework including physical implementation and block-level power-performance-area (PPA) for the first time. Contributions of the framework are listed below.

- The proposed framework includes the specially optimized power planning scheme, the dual-sided signal routing algorithm and the dual-sided RC extraction method.
- The power planning in the FFET is enabled by the novel Power Tap Cells which connect the frontside power rails to the BSPDN. The dual-sided signal routing is done through input pin redistribution and netlist partition without the need of bridging cells. RC extraction is achieved by merging the DEF files of both sides.
- For the first time, we verified that the FFET could outperform the CFET in terms of PPA at the block level. By optimizing the input pin density and BEOL routing layers, FFET with dual-sided signal routing delivers an extra 10.6% frequency gain without power degradation beyond the one with single-sided signal routing.

This paper is organized as follows. In Section II, we will detaily introduce the background of the FFET. Physical implementation and block-level PPA evaluation framework will be provided in Section III. Block-level PPA comparison between FFET and CFET and the optimization of input pin density together with BEOL routing layers will be given in Section IV, followed by conclusions in Section V.

## II. PRELIMINARIES

In this section, we will first introduce the 3D structure of the FFET standard cell, including 2 critical structures connecting dual-sided common signals. Then, the symmetric dual-sided cell design with the Split Gate (key design benefits of FFET) will be illustrated.

### A. Standard Cell Structure

With 3 or 2 M0 signal tracks and 1 shared M0 power rail placed on each side in a symmetric way, the FFET standard cells can be designed in 3.5T or 2.5T. In this work, we focus mainly on the 3.5T FFET with two-fin transistors (Fig. 2(a)) and assume the frontside transistor is nFET and the backside one is pFET. Under this setting, local power delivery is done through frontside VSS M0 power rails and backside VDD M0 power rails. Fig. 2(b) highlights 2 interconnect structures. One is the via called Gate Merge, which is used for merging the stacked nFET and pFET gates to form the n-p common gate in CMOS. The other one is the via called Drain Merge, which is used for the n-p common drain in CMOS.

### B. Symmytric Dual-sided Cell Design with Split Gate

Conventionally, intra-cell routing is implemented by frontside signal tracks. Though some design methods with backside intra-cell interconnects [4] were proposed recently, the backside routing resources are limited by the dominating backside-only power rails. Therefore, the backside intra-cell routing are supplementary to the frontside routing, with constrained routing capabilities. However, it is totally different in the FFET. The uniquely symmetric device structure paves the way for symmetric dual-sided intra-cell routing. This means the n-type logic in the CMOS can be realized solely by frontside signals and powers, and similar to the p-type logic on the backside. Benefited from it, there is no need for the frontside n-type logic to be routed on the backside or the p-type logic to be routed on the frontside, while in the CFET, the bottom p-type logic must be partly routed on the frontside. Thus, the supervias used in the CFET can be eliminated in the FFET, except for the Drain Merge which serves as the n-p sharing node in CMOS. Removing these supervias not only reduces FFET's footprint w.r.t the CFET (0.5T cell height reduction), but reduces intra-cell parasitics.

The separated frontside and backside gate process enables the Split Gate, which means the stacked nFET and pFET gates are controlled by 2 different signals and can be naturally realized by skipping the Gate Merge via. This is critical for the Transmission Gate and $C^2$MOS. For example, in the $C^2$MOS, some of the CMOS are controlled by a pair of complementary clock signals (Fig. 3(a)). With the help of the Split Gate, FFET can save more area compared to the CFET with common gate option only (Fig. 3(b-c)). The Split Gate is very area-efficient, especially in sequential logic like DFF for the Transmission Gates and $C^2$MOSs in it. Note that for fair standard cell area comparison, we choose 3.5T FFET and 4T CFET with the same active area footprint to assume the same intrinsic transistor characteristics. Fig. 4 shows the 3.5T FFET achieves area scaling of around 12.5% over the 4T CFET and gains extra area reduction in MUX and DFF due to the use of Split Gate explained previously. Note that, for comprehensive evaluation purpose, we have to admit that some of the FFET cells (e.g. AOI22 & OAI22) may have some area wasted due to extra Drain Merge needed in these cells. Fortunately, this degradation only occurs in very few cells in the whole library.

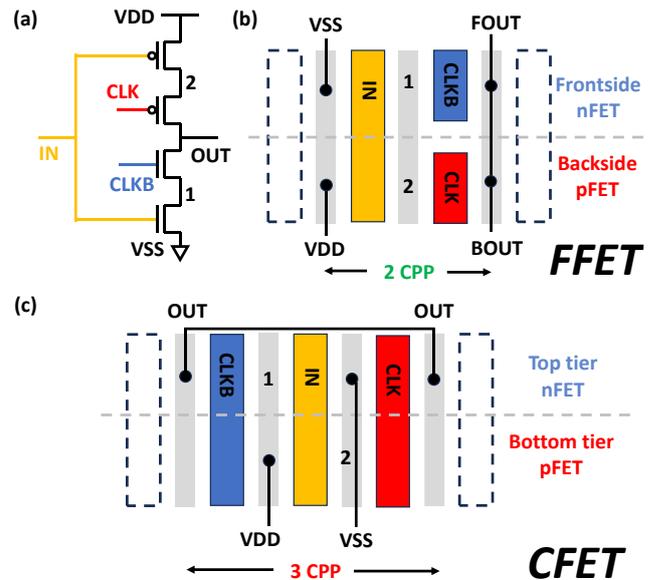

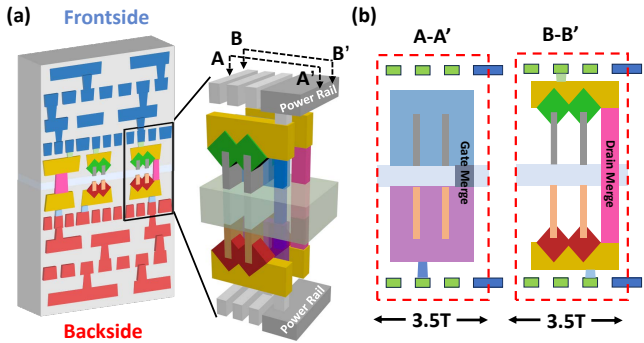

Fig. 2. (a) 3D structure of a 3.5T FFET standard cell with two-fin transistors, taking INVD1 as an example, based on [12]. (b) Cross-sections along the gate region (A-A') and source-drain region (B-B').

Fig. 3. (a) The schematic of the $C^2$MOS circuit, which is controlled by a pair of complementary clock signals (CLK & CLKB). (b) The vertical cross-section of FFET $C^2$MOS. The frontside gate (blue) and backside gate (red) in the Split Gate are controlled by the CLK and the CLKB respectively. (c) The vertical cross-section of CFET $C^2$MOS. The CFET wastes 1 more CPP because the CLKB cannot be stacked on the CLK without the Split Gate.

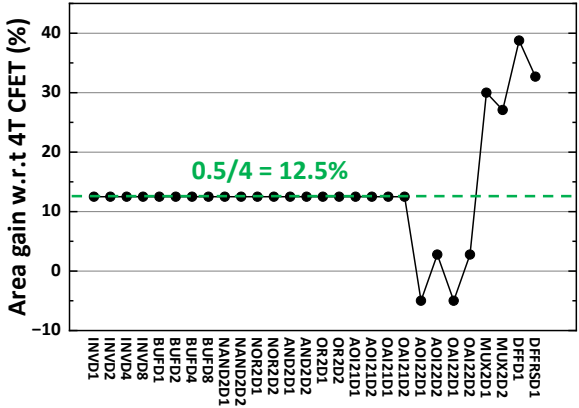

Fig. 4. Standard cell area comparison between 3.5T FFET and 4T CFET.

## III. METHODOLOGY

In this section, we will first give the principal and algorithm of dual-sided signal routing. Then we will discuss the powerplan based on the Power Tap Cell proposed. At last, we will introduce the whole physical implementation and PPA evaluation framework.

### A. Dual-sided Signal Routing

The principle of FFET's dual-sided routing is given as following. For logic gates, the output pin is typically composed of n-p common drains. In FFET cells, each output pin is made by the Drain Merge and gets connected to both frontside and backside M0 signal tracks. In this way, each output signal can be placed on the frontside (FOUT), the backside (BOUT), or both. Fig. 3(b) shows this ***dual-sided output pin***, taking C$^2$MOS as an example. With this dual-sided routing enabler, we can choose whether to use FOUT or BOUT freely for cascading based on which side the input pin of the next stage is located. Thanks to this, we can do the signal routing without using the bridging cells, as illustrated in Fig. 5. Furthermore, this interconnection method has no restriction in input pin location (frontside or backside) in standard cells. With this concept, we redesigned all the cells in Fig. 4 and found that for each cell, every input pin could be freely adjusted to the frontside or backside thanks to the enough resource of M0 signal tracks in 3.5T FFET.

Another possible solution is called the ***dual-sided input pin*** which means each input signal has its pins on both sides. The frontside input pins (FIN) and backside input pins (BIN) can be connected by the Gate Merge. Similarly, FIN or BIN can be chosen for cascading based on the location of the previous stage's output pins. However, the dual-sided input pins will lead to very high pin density and thus many cells cannot be achieved.

In conclusion, the ***dual-sided output pin*** is the only reasonable solution to the dual-sided signal routing in FFET technology.

We performed this novel dual-sided signal routing as described in Algorithm 1. Firstly, we set the decomposed frontside and backside nets according to the locations (frontside or backside) of the redistributed input pins on both sides. Their locations defined in the modified standard cell LEF files can be flexibly adjusted. (line 1) Secondly, we split the gate-level netlist into frontside nets and backside nets according to the result of the input pin redistribution. (line 2~8) Thirdly, the global & detailed routing are performed independently on both sides and two separate DEF files are generated by the tool. (line 9~10) Note that this dual-sided routing flow also supports bridging cells to transfer signals between the frontside and the backside.

**Algorithm 1**
**Input:** Dual-sided cell library *CL (in LEF)*, placed cells *C (in DEF)*, nets *N (in DEF)*.
**Output:** Routed dual-sided nets *T (in two DEFs)*.
1. Set the decomposed frontside nets *NF* and backside nets *NB*.
   # Decompose nets to dual-sided nets
2. For *n ∈ N*
3.    Initialize two nets *n.front* and *n.back* with *n.source*
4.    For *p ∈ n.sinks*
5.       Find the corresponding cell *c* of pin *p*
6.       Find the side type of *p* by the cell type in *CL* of c
7.       Assign *p* to *n.front* or *n.back* by side type.
8.    Add *n.front* to *NF* and add *n.back* to *NB*.
9. Perform routing for all nets in *NF* and *NB* independently.
10. Generate two DEFs to store routing results *T*.

### B. Power Planning

Practically, due to the chip package limitation, the power bump can only exist on one side of the chip. This means the power source of FFET can only be set on the backside because of the carrier wafer bonded to the FFET frontside in its process flow [12]. Therefore, the frontside VSS M0 power rails need special structures to be connected to the BSPDN. For here, we proposed a novel power planning scheme enabled by the so-called Power Tap Cell, inspired by a similar structure connecting BPR and the FSPDN in [1]. Fig. 6(a) shows the layout and cross section of FFET's PDN scheme. The backside VSS and VDD power stripes are arranged in an interleaved pattern. Backside M0 power rails are connected directly to the BSPDN while the frontside ones use the Power Tap Cells which are all placed right above the backside VSS power stripes. Fig. 6(b) shows the detailed intra-cell structure of the Power Tap Cell. For fair comparison, in the following sections, the 4T CFET with BPR also has BSPDN, enabled by the nTSV [1] as shown in Fig. 6(c).

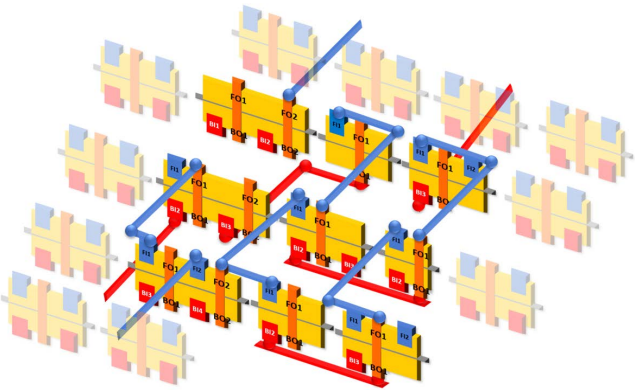

Fig. 5. 3D schematic demonstrating the dual-sided signal routing in the FFET. Frontside and backside routing can be performed independently.

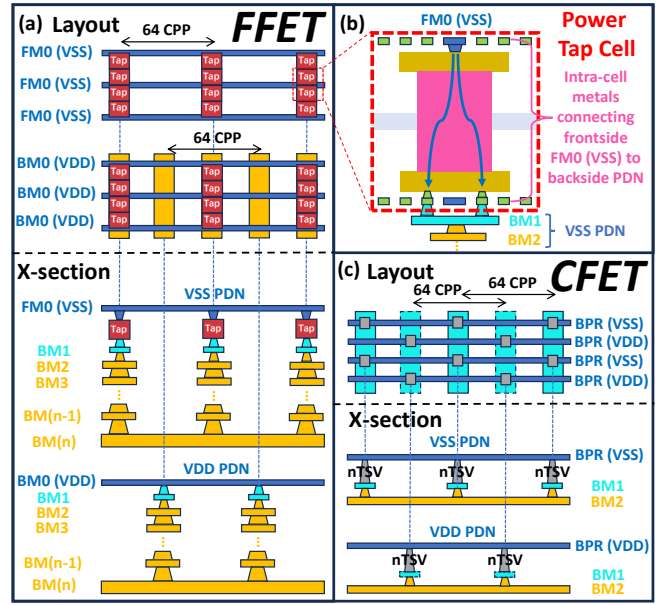

Fig. 6. (a) Layout and cross-section of FFET's VSS PDN and VDD PDN which are arranged in an interleaved pattern. The backside VDD power rails are connected to BSPDN directly. The frontside VSS power rails are connected to the backside VSS PDN by Power Tap Cells. (b) Detailed structure of the Power Tap Cell. By using intra-cell metals, the frontside VSS power rail is routed around the backside VDD power rail to connect to BSPDN. (c) BSPDN for CFET based on [1].

C. Design Framework

Our design framework for FFET is composed of two parts, i.e., physical implementation and power-performance assessment, as shown in Fig.7. The physical implementation includes floorplan, powerplan, placement, clock tree synthesis (CTS), and routing, where the powerplan and routing stages are novel and different from the conventional flow. The power-performance stage includes the novel RC extraction, which consists of DEF files merging, dual-sided RC extraction and static timing analysis.

In the physical implementation, we first defined the target area utilization and the aspect ratio of chip layout in the floorplan stage. In the powerplan stage, we implemented the novel power planning techniques as in Section III.B, including the BSPDN planning and the power tap cell placement. The techniques will occupy the user-defined layout area and metal wire space to ensure the power integrity and the even distribution of power supply across both sides of the chip. By these steps, we achieved a complete BSPDN for FFET-based design. The following stage is the placement, including the standard cell & IP placement and IO planning. In the placement stage, the tool cannot move the placed power tap cells and need to optimize the routability of designs according to the powerplan result. Subsequently, the CTS stage is performed, which is the same as the conventional flow. The dual-sided signal routing is the last critical stage to accomplish all wire connections on both sides, as already discussed in Section III.A. The power-performance evaluation is then conducted to analyze the two DEF files of both sides, which stores the whole block's physical information. To adapt this to standard tools, we first merged the two DEFs into one DEF. It contains the P&R information of all the frontside and backside layers and is used in the accurate dual-sided RC extraction. In the end, power and achieved frequency is analyzed by commercially available tools based on the RC net of the block.

IV. EXPERIMENTAL RESULTS

To compare the PPA between the FFET and the CFET at the block level, we first characterized the standard cell libraries listed in Fig. 4 based on a virtual 5nm node PDK. For fair comparison, we assume the 3.5 FFET and 4T CFET have the same two-fin transistor structure and the same intrinsic transistor characteristics. Table I gives part of the results of the library characterization. 3.5T FFET outperforms the 4T CFET substantially in terms of timing and exhibits better power efficiency. This is consistent with the smaller parasitics in the FFET as mentioned in Section II.B. To reduce the synthesis complexity, we assume the characteristics of the same cell remains the same across different input pin configurations, considering that their structure differences mainly exists in the M0 layer but intra-cell parasitics are barely affected by these structural differences in the M0 layer.

TABLE I. PART OF LIBRARY CHARACTERIZATION RESULTS. INV AND BUF CELLS WITH DRIVING CAPABILITIES OF D1/D2/D4 WERE SELECTED.

| Benchmark | KPI Diff of FFET Libraries w.r.t CFET | | | | | |
|---|---|---|---|---|---|---|
| | *INVD1* | *INVD2* | *INVD4* | *BUFD1* | *BUFD2* | *BUFD4* |
| Transition power[a] | +0.3% | +0.3% | +0.2% | -3.0% | -10.9% | -11.8% |
| Leakage power[b] | 0.0% | 0.0% | 0.0% | 0.0% | 0.0% | 0.0% |
| Rise timing | -2.5% | -2.8% | 6.8% | -10.1% | -12.8% | -13.6% |
| Fall timing | -8.1% | -9.9% | -13.6% | -10.7% | -14.4% | -15.8% |
| Rise transition | -1.1% | -1.2% | -4.9% | -3.9% | -8.4% | 9.2% |
| Fall transition | -4.0% | -2.4% | -3.4% | -5.1% | -6.5% | -9.7% |

[a.] Transition power = Rise power + Fall power.
[b.] The leakage power is determined by the intrinsic transistor characteristics, thus being the same.

A 32-bit RISC-V core was used as the benchmark design. P&R was realized by Cadence® Innovus™ Implementation System and RC extraction was implemented by Synopsys® StarRC™. To minimize the area cost, we did not use the bridging cells. Design rules including frontside and backside BEOL routing layers are listed in Table II. Both FFET and CFET are powered by BSPDN as mentioned in Section III.B, with 64 CPP of power stripe pitch. For the CFET, the highest PDN layer is BM2. While, for the FFET, it is determined by the highest signal routing layer on the backside. Note that, the "routing" in this evaluation refers exclusively to the inter-cell routing described in Section IV. FM0 and BM0 are only used for intra-cell routing, so they are not included in the routing layers. Moreover, we assume that a P&R result is valid only if the total design rule violation number is below 10.

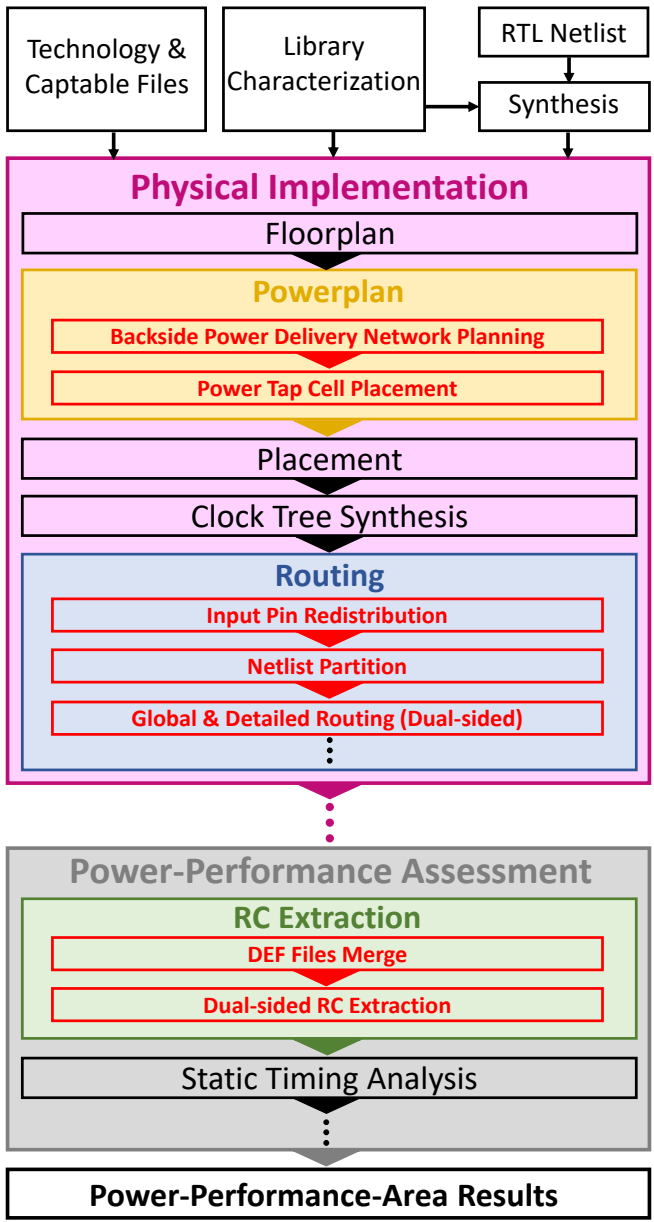

Fig. 7. The overview of the physical implementation and block-level PPA evaluation framework for the FFET.

TABLE II. DESIGN RULES INCLUDING FRONTSIDE AND BACKSIDE BEOL METAL LAYERS BASED ON A VIRTUAL 5NM NODE PDK.

| Layers | Pitch (nm) | |
| --- | --- | --- |
| | 4T CFET | 3.5T FFET |
| FM12 | 720 | 720 |
| FM11 | 126 | 126 |
| FM10-5 | 76 | 76 |
| FM4-3 | 42 | 42 |
| FM2 | 30 | 30 |
| FM1 | 34 | 34 |
| FM0 | 28 | 28 |
| Poly | 50 | 50 |
| BPR | 120 | / |
| BM0 | / | 28 |
| BM1[c] | 3200 | 34 |
| BM2[c] | 2400 | 30 |
| BM3-BM4 | / | 42 |
| BM5-BM10 | / | 76 |
| BM11 | / | 126 |
| BM12 | / | 720 |

[c]. In the CFET, BM1 and BM2 are only used for PDN, thus with larger metal pitches.

To investigate the maximum advantages provided by dual-sided signal routing in enhancing FFET routability, we first made the input pins evenly distributed on both sides and use routing layers up to FM12 and BM12 (FFET $FM_{12}BM_{12}$). At synthesis target frequency = 1.5 GHz, FFET $FM_{12}BM_{12}$ achieves maximum utilization of 86% which is higher than the CFET as shown in Fig. 8(a). Actually, 86% utilization is almost the upper limit, because utilization above 86% results in placement violations between standard cells and Power Tap Cells considering the designed 64 CPP of power stripe pitch. 25.1% post-P&R core area reduction at respective minimum core area is achieved, and 23.3% reduction at the same utilization which is validated by the core layout comparison in Fig. 8(b). For fairer comparison, FFET with frontside-only routing layers up to FM12 (FFET $FM_{12}$) was also physically implemented as shown in Fig. 8(c). Without the backside signal routing, the maximum utilization reduces to 76% and area gain reduces to 15.4% at respective minimum core area. This routability degradation comes from the higher pin density in FFET $FM_{12}$ with frontside-only signals (pins) compared with the CFET due to FFET's smaller cell area.

Next, we compared the PPA between 3.5T FFET $FM_{12}$ and 4T CFET, both with single-sided signal routing. We swept the synthesis target frequency from 500 MHz to 3 GHz to get a wide range of power-frequency relationship. Fig. 9 gives that the 3.5T FFET $FM_{12}$ outperforms the 4T CFET by 25% in frequency and 11.9% in power at the same 76% utilization. Besides, the frequency-area relationship is shown in Fig. 10. At 1.5 GHz target frequency, the FFET $FM_{12}$ attends a 16.0% higher frequency compared to the CFET's maximum frequency with the same core area and reached a 23.4% higher frequency at respective maximum frequency. These all prove that FFET's PPA advantages against the CFET at the cell level continue to persist at the block level even without backside pins.

Additionally, we explored the design space of dual-sided signal routing by tunning the input pin density and BEOL routing layer number, taking the FFET $FM_{12}$ as the baseline. 5 types of design of experiments (DoEs) with various backside input pin density ratios ranging from 4% to 50% were implemented. Fig. 11 shows the power-frequency relationship among these 5 designs. The data points were extracted by sweeping the utilization from 46% to 76% at the same 1.5 GHz target frequency, all with the same $FM_{12}BM_{12}$ routing layer pattern. The FFET $FP_{0.5}BP_{0.5}$ (this means 50% input pins are on the frontside and 50 % input pins ar e on the backside, etc.) and the FFET $FP_{0.6}BP_{0.4}$ exhibits the best power-frequency characteristics, followed by the FFET $FP_{0.7}BP_{0.3}$, with FFET $FP_{0.84}BP_{0.16}$ and FFET $FP_{0.96}BP_{0.04}$ trailing behind.

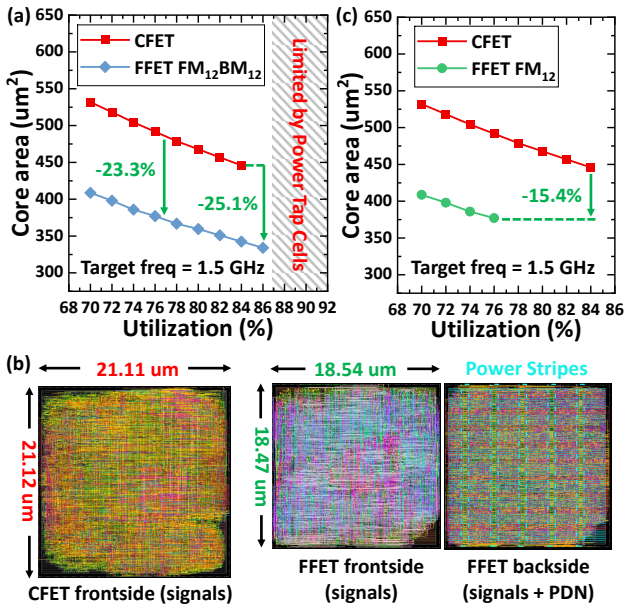

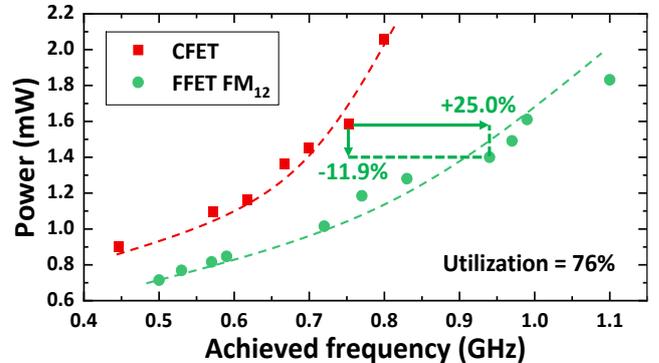

Fig. 9. Power-frequency relationship comparison between the CFET and FFET $FM_{12}$. The synthesis target frequency varies from 500 MHz to 3 GHz.

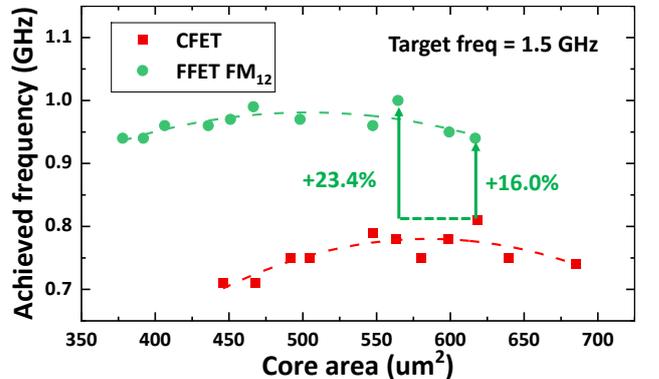

Fig. 10. Frequency-area relationship comparison between the CFET and FFET $FM_{12}$. The synthesis target frequency is 1.5 GHz.

Fig. 8. (a) Core area vs utilization comparing CFET and FFET $FM_{12}BM_{12}$. Maximum utilization is limited by the placement of the Power Tap Cells under power stripes. (b) Core layout comparison at 84% utilization. CFET backside layout with BSPDN composed of BM1 & BM2 is not shown here. FFET backside layout shows the power stripes over the Power Tap Cell. (c) Core area vs utilization comparing the CFET and the FFET $FM_{12}$.

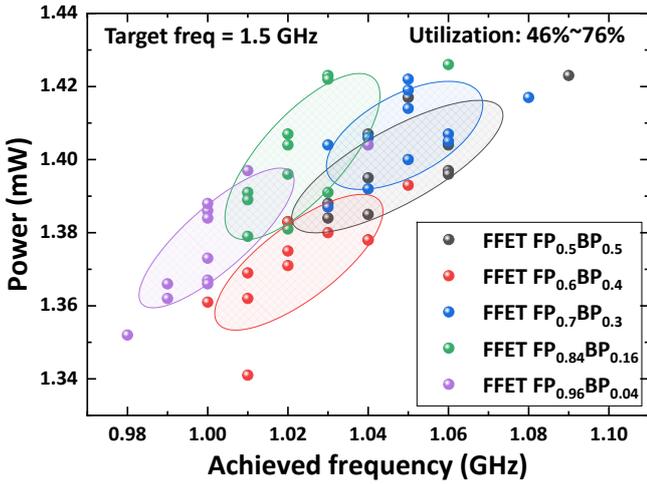

Fig. 11. Power-frequency relationship comparison among 5 types of DoEs with different input pin density and the same $FM_{12}BM_{12}$ routing layer pattern. Confidence ellipses were used to illustrate the distribution range of power and frequency. For each ellipse, the confidence level is set to be 50%.

However, with too many metal layers used, $FM_{12}BM_{12}$ faces many challenges and is costly in practical manufacturing processes. Therefore, in each input pin density pattern, we limited the total routing layer number to 12 and attempted all possible combinations of frontside and backside routing layers that could remain the routability. The experimental results of input pin density and BEOL routing layer number co-optimization are listed in Table III. Compared with the baseline FFET FM12, FFET $FP_{0.5}BP_{0.5}$ routed with $FM_6BM_6$ achieves the biggest frequency gain of 10.6% in all DoEs without power degradation and FFET $FP_{0.7}BP_{0.3}$ routed with $FM_8BM_4$ or $FM_7BM_5$ delivers the highest frequency improvement, with a gain of 12.8%, despite a minor power degradation of 1.4%.

TABLE III. POWER-FREQUENCY VARIATION WITH DIFFERENT INPUT PIN DENSITY AND ROUTING LAYERS COMPARED WITH THE BASELINE FFET

| Input Pin Density | Routing Layer Pattern | Achieved Frequency Diff | Power Diff |
|---|---|---|---|
| $FP_{0.96}BP_{0.04}$ | $FM_{10}BM_2$ | +5.3% | -2.9% |
|  | $FM_9BM_3$ | +5.3% | -2.1% |
| $FP_{0.84}BP_{0.16}$ | $FM_9BM_3$ | +8.5% | -0.7% |
|  | $FM_8BM_4$ | +9.6% | +0.7% |
| $FP_{0.7}BP_{0.3}$ | $FM_9BM_3$ | +8.5% | -2% |
|  | **$FM_8BM_4$** | **+12.8%** | **+1.4%** |
|  | **$FM_7BM_5$** | **+12.8%** | **+1.4%** |
| $FP_{0.6}BP_{0.4}$ | $FM_8BM_4$ | +6.3% | -4.3% |
|  | $FM_7BM_3$ | +8.5% | -2.9% |
|  | $FM_6BM_6$ | +7.4% | -3.6% |
| $FP_{0.5}BP_{0.5}$ | $FM_8BM_4$ | +9.6% | -1.4% |
|  | $FM_7BM_3$ | +10.6% | -0.7% |
|  | **$FM_6BM_6$** | **+10.6%** | **-1.4%** |

To further decrease the BEOL layer cost in FFET, we reduced the number of routing layers on both sides simultaneously in FFET $FP_{0.5}BP_{0.5}$. Fig. 12 shows that the maximum utilization remains a constant value of 86% until the number of routing layers drops below 4 on each side. This also suggests that the core area scaling is limited by the Power Tap Cell, not the routability as long as the routing layer number is no less than 4 on each side. It also worths mentioning that the utilization can still reach 70% with only 2 routing layers on both frontside and backside. Fig. 13 shows the power efficiency variation in the FFET $FP_{0.5}BP_{0.5}$ routed with routing layer number reduced from 12 to 3 on each side at 1.5 GHz target frequency and 76% utilization. Power efficiency of FFET degrades by only 0.68% even when the number of routing layers is reduced from 12 to 5 on each side. In conclusion, the trends in routability and power efficiency validate the great space for cost-friendly design in the FFET architecture.

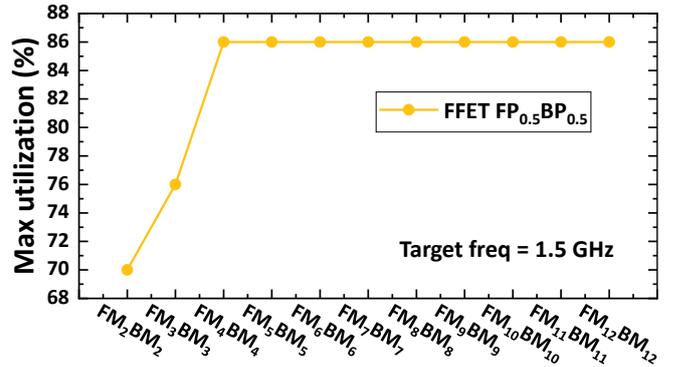

Fig. 12. Maximum utilization of FFET $FP_{0.5}BP_{0.5}$ varying the number of frontside and backside routing layers simultaneously.

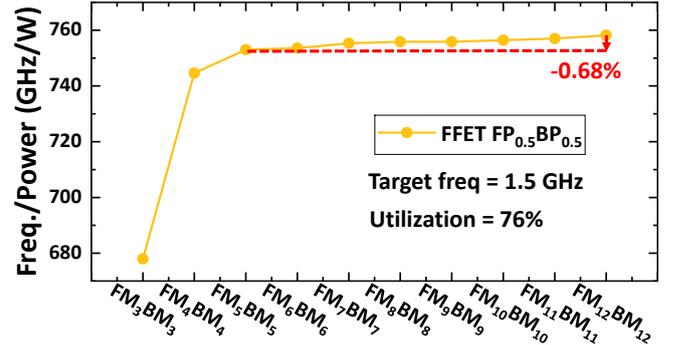

Fig. 13. The variation of the power efficiency of FFET $FP_{0.5}BP_{0.5}$ w.r.t the number of frontside and backside routing layers.

## V. CONCLUSIONS

The Flip FET is a novel stacked transistor technology combining the backside 3D transistor stacking, backside power and backside signals. It's a first of a kind technology with fully functional backside. In this work, we proposed a full-spectrum physical implementation and PPA evaluation framework on the Flip FET for the first time. Power delivery network was built through the placement of the Power Tap Cell during the powerplan. Dual-sided signal routing with dual-sided pins in standard cells was also successfully implemented. Furthermore, dual-sided RC net was extracted for precise power-frequency analysis. We have proved that FFET features greater scalability in terms of PPA beyond the CFET. Thanks to the flexibility of input pin density redistribution and dual-sided signaling with various BEOL routing layer options, FFET can be further improved in routability and PPA while keeping the manufactural cost reasonable. This work demonstrates that the Flip FET has great potential as a key enabler for the continuous scaling and advancement of high performance computing in the future.

ACKNOWLEDGMENTS

This work was supported in part by the National Key Research and Development Program of China under Grant 2023YFB4402200; in part by the 111 Project under Grant QYJS-2023-2301-B; and in part by the National Natural Science Foundation of China under Grant 92464206.

## References

[1] R. Chen et al., "Power, Performance, Area and Thermal Analysis of 2D and 3D ICs at A14 Node Designed with Back-side Power Delivery Network," 2022 International Electron Devices Meeting (IEDM), San Francisco, CA, USA, 2022, pp. 23.4.1-23.4.4.

[2] G. Sisto et al., "Block-level Evaluation and Optimization of Backside PDN for High-Performance Computing at the A14 node," 2023 IEEE Symposium on VLSI Technology and Circuits (VLSI Technology and Circuits), Kyoto, Japan, 2023, pp. 1-2.

[3] A. Veloso et al., "Backside Power Delivery: Game Changer and Key Enabler of Advanced Logic Scaling and New STCO Opportunities," 2023 International Electron Devices Meeting (IEDM), San Francisco, C-A, USA, 2023, pp. 1-4.

[4] B. -S. Kim et al., "Expanding Design Technology Co-Optimization Potentials with Back-Side Interconnect Innovation," 2024 IEEE Symposium on VLSI Technology and Circuits (VLSI Technology and Circuits), Honolulu, HI, USA, 2024, pp. 1-2.

[5] P. Vanna-iampikul et al., "Back-side Design Methodology for Power Delivery Network and Clock Routing," 2024 IEEE Symposium on VLSI Technology and Circuits (VLSI Technology and Circuits), Honolulu, HI, USA, 2024, pp. 1-2.

[6] R. Chen et al., "Design and Optimization of SRAM Macro and Logic Using Backside Interconnects at 2nm node," 2021 IEEE International Electron Devices Meeting (IEDM), San Francisco, CA, USA, 2021, pp. 22.4.1-22.4.4.

[7] T. -C. Lin, F. -Y. Hsu, W. -K. Mak and T. -C. Wang, "An Effective Netlist Planning Approach for Double-sided Signal Routing," 2024 29th Asia and South Pacific Design Automation Conference (ASP-DAC), Incheon, Korea, Republic of, 2024, pp. 288-293.

[8] J. Ryckaert et al., "The Complementary FET (CFET) for CMOS scaling beyond N3," 2018 IEEE Symposium on VLSI Technology, Honolulu, HI, USA, 2018, pp. 141-142.

[9] C. -K. Cheng, C. -T. Ho, D. Lee and D. Park, "A Routability-Driven Complimentary-FET (CFET) Standard Cell Synthesis Framework using SMT," 2020 IEEE/ACM International Conference On Computer Aided Design (ICCAD), San Diego, CA, USA, 2020, pp. 1-8.

[10] S. M. Y. Sherazi et al., "CFET standard-cell design down to 3Track height for node 3nm and below," Proc. SPIE 10962, Design-Process-Technology Co-optimization for Manufacturability XIII, 1096206 (20 March 2019).

[11] O. Zografos et al., "Design enablement of CFET devices for sub-2nm CMOS nodes," 2022 Design, Automation & Test in Europe Conference & Exhibition (DATE), Antwerp, Belgium, 2022, pp. 29-33.

[12] H. Lu et al., "First Experimental Demonstration of Self-Aligned Flip FET (FFET): A Breakthrough Stacked Transistor Technology with 2.5T Design, Dual-Side Active and Interconnects," 2024 IEEE Symposium on VLSI Technology and Circuits (VLSI Technology and Circuits), Honolulu, HI, USA, 2024, pp. 1-2.